\documentclass{article}

\usepackage{arxiv}

\usepackage[utf8]{inputenc} 
\usepackage[T1]{fontenc}    
\usepackage{hyperref}       
\usepackage{url}            
\usepackage{booktabs}       
\usepackage{amsfonts}       
\usepackage{nicefrac}       
\usepackage{microtype}      
\usepackage{graphicx}
\usepackage[numbers]{natbib}
\usepackage{doi}

\title{A Survey of Security Research for Operating Systems}

\author{
	\href{https://orcid.org/0000-0001-5596-282X}{\includegraphics[scale=0.06]{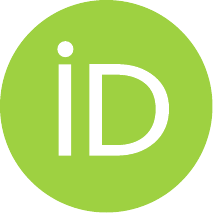}\hspace{1mm}Masaki Hashimoto} \\
	Graduate School of Information Security \\
	Institute of Information Security\\
	\texttt{hashimoto@iisec.ac.jp} \\
	\And
	Ruo Ando \\
	Institute of Information Security \\
	National Institute of Information and \\ Communications Technology (NICT) \\
	\And
	Toshiyuki Maeda \\
	Graduate School of Information \\ Science and Technology \\
	The University of Tokyo \\
	\And
	Hidehiko Tanaka \\
	Graduate School of Information Security \\
	Institute of Information Security \\
}

\renewcommand{\shorttitle}{A Survey of Security Research for Operating Systems}

\hypersetup{
pdftitle={A Survey of Security Research for Operating Systems},
pdfauthor={Masaki Hashimoto, Ruo Ando, Toshiyuki Maeda, Hidehiko Tanaka},
pdfkeywords={Operating system security, Virtualization, Formal verification, Access control, Reference monitor, Trusted Computing Base},
}

\begin{document}
\maketitle

\let\thefootnote\relax\footnotetext{
	This document is an English translation of the authors' paper originally published in Japanese: M.~Hashimoto, R.~Ando, T.~Maeda, and H.~Tanaka, ``A Survey of Security Research for Operating Systems,'' IPSJ Journal: Computing Systems (ACS), Vol.~5, No.~2, pp.~51--62 (Mar.~2012).
	\copyright~Information Processing Society of Japan (IPSJ).
}

\begin{abstract}
	In recent years, information systems have become the social infrastructure, so that their
	security must be improved urgently. In this paper, we introduce the results of the survey of
	virtualization, operating system verification and access control technologies in association
	with the design requirements of the reference monitor. Additionally, we show the prospects and
	challenges for each technology.
\end{abstract}

\keywords{Operating system security \and Virtualization \and Formal verification \and Access control \and Reference monitor \and Trusted Computing Base}

\section{Introduction}
\label{sec:introduction}
In recent years, because information systems have become a social infrastructure, strengthening their security has become an important issue. Our social life today is established by various forms of support provided by information systems, so the damage caused when a security incident occurs has become more serious than ever. At the same time, because the demands placed on information systems have diversified along with the expansion of their applications, their configurations have grown complex; together with the explosive increase in the number of nodes connected to the Internet, this has become one factor behind the year-on-year increase in the number of security incidents.

To strengthen information security, fundamental support from technology, management, legislation, and ethics is indispensable, and a wide variety of research, development, and study is being carried out toward this end. Focusing on technology in particular, a broad range of research is conducted---cryptography, intrusion detection, authentication, evidence preservation, and so on---but almost all of these assume the soundness of the foundation on which their functions operate. That foundation is the so-called operating system (OS). If the OS is vulnerable, the various security-strengthening functions running on top of it are castles built on sand; and even when the OS is sound, what kind of security-support functions the OS should provide to the upper layers remains an issue that requires study.

For this reason, the security strengthening of the OS itself and the security-support functions that a sound OS should provide have been studied for a long time. Beginning with the U.S.\ Department of Defense's Trusted OS, and including more recent examples such as SELinux and Trusted Solaris, many research and implementation examples exist. However, although these are important elements that fundamentally reduce various threats, they cannot yet be called sufficient solutions, and active research continues in many places even today.

In this paper, we position the OS as the most basic software that guarantees information security, and we introduce recent research trends focusing in particular on OS virtualization technology, OS program verification technology, and access control technology. First, in Section~\ref{sec:overview}, we explain the classification policy of this paper for security research on OSs in relation to the design requirements of the reference monitor, and within this we describe the relationship among the three technologies above. Then, in Sections~\ref{sec:virtualization}, \ref{sec:verification}, and \ref{sec:access_control}, we introduce the research trends of each and describe future challenges. Finally, in Section~\ref{sec:conclusion} we summarize this paper.

\section{Security Research in Operating Systems}
\label{sec:overview}
The Trusted Computing Base (TCB) is a concept defined by the U.S.\ Department of Defense in the TCSEC (Trusted Computer System Evaluation Criteria)~\cite{Orange:1983}, and it is the basic component for realizing a high-assurance system. The TCB includes a reference monitor\footnote{The reference monitor is a mechanism that enforces legitimate access relationships between access subjects and access targets; it was presented as an essential element of secure computing by the Anderson Report~\cite{anderson:1972}, published in 1972.} and a formally verifiable security policy. For the reference monitor in particular, its design requirements are given as: (i) being tamper-resistant, (ii) being impossible for the controlled targets to bypass, and (iii) being small enough to guarantee completeness.

Because the OS sits at the lowest layer of the software stack composing an information system, it can capture the behavior of the upper layers, and thus naturally satisfies design requirement (ii) of the reference monitor. For this reason, the OS has traditionally played the role of the main software composing the TCB, and research that guarantees information security by constructing the OS more robustly as a reference monitor continues today. At the same time, because the OS also plays the role of basic software providing an execution environment to application programs, research from aspects other than that of the reference monitor has also progressed, in order to respond to the evolution of hardware and to new demands on information systems. In particular, in recent years, with the spread of distributed computing represented by embedded systems, mobile devices, and the cloud, research on new OS technologies that respond to demands from these areas is being conducted in many places.

Therefore, OS research can be viewed from various aspects; but in this paper, particularly in relation to information security, we organize recent OS research into the following three classifications. That is, among the threats to information systems, we focus on those deeply related to the OS---attacks on the OS itself and attacks on programs running on the OS. In this classification, we organize OS verification technology mainly as a countermeasure against the former, access control technology mainly as a countermeasure against the latter, and virtualization technology as a countermeasure against both. The encryption of data and I/O, authentication technology, digital forensics, and the like are also technologies related to the OS and information security, but they are outside the scope of this paper.

\begin{enumerate}
\item Virtualization technology

Virtualization technology has been studied for a long time in order to use hardware resources effectively, and in recent years it has become widely used as a foundational technology of cloud computing. When OS virtualization technology is viewed particularly in relation to the TCB, it relates to all of the reference monitor requirements (i)--(iii).

\item OS verification technology

OS verification technology confirms the integrity of the OS by applying conventional program verification techniques to the OS. As described above, being able to guarantee the integrity of the TCB is a requirement for realizing a high-assurance system, and virtualization technology and OS verification technology relate to reference monitor requirement (iii).

\item Access control technology

Access control technology guarantees the safety of the entire information system using a reference monitor, and relates to reference monitor requirement (ii). It assumes that the reference monitor is complete, and its main challenge is how to control other programs.

\end{enumerate}

\section{OS Virtualization Technology}
\label{sec:virtualization}
In recent years, due to the remarkable improvement in hardware performance, research and development on OS virtualization technology has become active. OS virtualization technology is also widely used as a foundational technology of cloud computing, and strengthening its security has become an important issue. In this section, we organize and introduce recent research on virtualization technology from four aspects: ``observation of virtual machines by the hypervisor,'' ``virtualization of main memory,'' ``virtualization of the I/O mechanism,'' and ``integrity verification of virtual machines.'' We also describe future challenges.

Regarding the rationale for the four classifications: first, in order to apply appropriate information-security measures, it is important that, for a certain event occurring inside an information system, the event be observable from the defender's side and unobservable from the attacker's side; considering that research toward this is being actively conducted, we set the observation of virtual machines by the hypervisor as the first item. Next, because the current targets of observation by the hypervisor are mainly main memory and I/O, we organized each of these as well. The final item is organized separately because, in order to exclude software introduced by the attacker for the purpose of observation, integrity verification of the execution sequence including the boot process is necessary, and research premised on hardware support is being conducted in many places toward this end.

\subsection{Observation of Virtual Machines by the Hypervisor}
\label{sec:vm_observation}
The hypervisor is located between the virtual machine and the hardware, and performs resource management and scheduling for the virtual machines. The benefits enjoyed through the hypervisor are, first, on the operational side, that multiple virtual machines can be centrally managed---that is, the resources allocated can be dynamically expanded or contracted according to operating conditions. The second is security strengthening of virtual machines: by using the hypervisor, one can strengthen access control on general-purpose OSs that have become complex and contain vulnerabilities, and one can construct observation and analysis units that cannot be detected by malware. Representative implementation examples of hypervisors include Xen~\cite{Barham:2003:XAV:945445.945462}, KVM~\cite{kivity2007kvm}, and VMware~\cite{Waldspurger:2002:MRM:844128.844146}.

The concept of monitoring a virtual machine using a hypervisor and performing processing such as observation, defense, and isolation was presented as VMI (Virtual Machine Introspection)~\cite{garfinkel2003virtual}. There, VMI is pointed out to be effective for unauthorized-access detection and defense on three points: that the hypervisor-side code cannot be modified from the virtual machine, that all states of the virtual machine can be observed from the hypervisor side, and that code issued from the virtual machine can be captured. Many papers on virtual-machine observation that reference this work have since been published.

VMI observation methods include active methods and passive methods. The active method acquires the state of the virtual machine from the outside at fixed intervals and extracts information; a representative research example is Volatility~\cite{Volatility}, which acquires and analyzes memory snapshots. The passive method draws out related information when an event such as a resource access occurs inside the virtual machine; representative research examples include Lares~\cite{10.1109/SP.2008.24}, Xenprobes~\cite{Quynh:2007:XLU:1364385.1364387}, and VMScope~\cite{Breslau:2006:VVM:1162638.1162648}.

As concrete implementation methods of the passive and active methods, there are two approaches: the in-the-box (internal observation) method and the out-of-the-box (external observation) method. These are often discussed in relation to the detectability of the observation mechanism from the attacker's side, the resolvability of the semantic gap~\cite{Chen:2001:VBR:874075.876409}\footnote{The semantic gap refers to the fact that the semantic information related to resource accesses inside a virtual machine cannot be reconstructed from the information obtainable on the hypervisor side. That is, resolving the semantic gap means making it possible for the hypervisor to reconstruct the semantics of resource accesses occurring on the virtual machine.}, and the trade-off in implementation cost~\cite{Sharif:2009:SIM:1653662.1653720}. At present, because it is difficult for the observation mechanism to be detected from inside the observed virtual machine, the external observation method is often adopted. On the other hand, the external observation method has the issue that the amount of observable information is reduced compared with the internal observation method.

\subsection{Virtualization of Main Memory}
\label{sec:memory_virtualization}
When implementing virtualization technology, the handling of main memory, particularly the paging mechanism, becomes an important issue. The physical address used by a virtual machine is a pseudo-physical address virtualized by the hypervisor, and in order to arbitrate access to the true physical address, a double address translation is required compared with the ordinary case. This two-or-more-stage paging peculiar to virtualization technology may be performed in software or in hardware.

A representative example of software-based main-memory virtualization is Xen's Shadow Paging. Shadow Paging captures page faults in main memory and virtualizes the virtual machine's accesses to main memory; the mechanism that translates the physical address recognized by the virtual machine into the physical address of the real machine is called the Shadow Page Table. Related to this, research that detects malicious kernel extensions such as rootkits by modifying the Shadow Page Table is being actively conducted.

For example, Panorama~\cite{Yin:2007:PCS:1315245.1315261} proposes a taint-detection method using Google Desktop~\cite{googledesktop} as a case study, and Ether~\cite{Dinaburg:2008:EMA:1455770.1455779} proposes a method for detecting malware using hardware virtualization-support features. Similarly, nEther~\cite{Pek:2011:NID:1972551.1972554} proposes a method for detecting, from the guest OS, malware running externally, by collecting data from the RDTSC.

On the other hand, from the attacker's viewpoint, in 2006 SubVirt~\cite{10.1109/SP.2006.38} and Blue Pill~\cite{rutkowska2006subverting}---which can construct rootkits invisible from the guest OS by exploiting the powerful isolation provided by the hypervisor---were proposed. Similarly, Ristenpart et al.\ showed that, targeting Amazon EC2, a side-channel attack between virtual machines running on the same physical machine is possible~\cite{Ristenpart:2009:HYG:1653662.1653687}. In addition, Chen et al.\ comprehensively surveyed the methods by which malware detects analysis by the defending hypervisor, and classified them~\cite{chen2008towards}.

Representative examples of hardware-based main-memory virtualization include Intel VT-d and AMD-V. In Intel VT-d and AMD-V, a hardware memory-management mechanism is available for virtualization technology, which can reduce the load of address translation and the implementation load on the hypervisor. Related to this, HUKO~\cite{xiong2011practical} is research that detects unauthorized kernel extensions using Intel VT-d's EPT (Extended Page Table). HUKO physically isolates the table that was shared under Shadow Paging and gives each guest OS its own, thereby reducing the paging overhead of the guest OS while strengthening access control. Regarding protection by address-space separation using virtualization technology, there is also SIM~\cite{Sharif:2009:SIM:1653662.1653720}; whereas SIM uses Shadow Paging, HUKO uses hardware-assisted paging to reduce the load on VMM transitions and the TLB.

\subsection{Virtualization of the I/O Mechanism}
\label{sec:io_virtualization}
When the hypervisor performs security strengthening such as access control or integrity verification on virtual machines, in addition to the main-memory virtualization introduced in the previous section, it is necessary to virtualize I/O. Like main-memory virtualization, I/O virtualization has software-based and hardware-based variants.

In software-based I/O virtualization, the capture and relay of I/O requests is performed by device drivers. Xen's Split Kernel Driver is a representative example; it places shared memory between the virtual machine and the hypervisor and virtualizes I/O using event channels. As an application of this, XenAccess~\cite{payne2007secure} uses the blktap driver, a Split Kernel Driver, to modify virtual-machine events and reduce the load on the whole system. Also, sHype~\cite{sailer2005shype} can enforce behavior according to a security policy at the time of virtual-machine snapshot creation and destruction, or when a virtual interrupt occurs.

Another implementation example is BitVisor~\cite{Shinagawa:2009:BTH:1508293.1508311}. BitVisor is a hypervisor that adopts an architecture called para-passthrough. Compared with other hypervisors, BitVisor captures, among the I/O requests generated by the virtual machine, only those related to access control or encryption, and strengthens security by performing intermediate processing on them. By adopting this architecture, the hypervisor side need only process control I/O and data I/O, and protection between virtual machines and scheduling become unnecessary. BitVisor has succeeded in keeping the code size to about 20\,KLOC in the core part, and the para-passthrough driver to about one-tenth of an ordinary driver. From the viewpoint of the CIA (confidentiality, integrity, availability) of information security, whereas other hypervisors are devoted to ensuring confidentiality, BitVisor is characterized by having functions effective not only for confidentiality but also for ensuring integrity.

In hardware-based I/O virtualization, the IOMMU, which supports direct access to hardware by virtual machines, has been proposed. The IOMMU realizes the address-remapping function during DMA (Direct Memory Access) in hardware, making it possible for a virtual machine to directly handle the address of a physical device. Related to this, in Intel VT-d, a function that prevents unauthorized code transfer using DMA is further realized by TXT (Trusted Execution Technology)~\cite{inteltxt}. TXT intervenes in the address-remapping function that occurs during DMA and prohibits DMA to protected regions. Also, by Intel VT-i and Intel VT-d, complete separation of the I/O space including the memory accesses of the guest OS becomes possible, which can ensure confidentiality.

\subsection{Integrity Verification of Virtual Machines}
\label{sec:vm_integrity}
As attacks on virtual machines increase, research that performs integrity verification on the components of the virtualized machine has been actively conducted. A device that enables integrity verification of a physical machine is the TPM (Trusted Platform Module); the vTPM~\cite{Berger:2006:VVT:1267336.1267357}, which virtualizes this TPM and performs attestation of multiple virtual machines on a physical machine, has been proposed.

In the vTPM, in order to capture the I/O requests from the virtual machine to the TPM, the Split Device Driver introduced earlier is implemented on both the virtual machine and the virtual machine monitor. Trusted Boot using the TPM comes in two kinds: SRTM (Static Root of Trust Measurement) and DRTM (Dynamic Root of Trust Management). SRTM is used in Sailer et al.'s research~\cite{Sailer:2004:DIT:1251375.1251391} and in BitLocker~\cite{bitrocker}. In Sailer et al.'s 2004 research, a method for ensuring a chain of trust from the boot loader through the kernel to the executed applications is proposed. By this, the code and data loaded into the OS come to have their integrity dynamically verified by the TPM. Kauer proposed attack methods against SRTM as of 2007 and a method for constructing the Open Secure Loader (OSLO) using AMD's SKINIT~\cite{Kauer:2007:OIS:1362903.1362919}.

Also, as a more advanced integrity-verification method, HIMA~\cite{azab2009hima}, which can verify the consistency of TOCTTOU (time of check to time of use), has been proposed. In HIMA, as design goals, priority is placed on strong isolation and on defense against TOCTTOU attacks; as methods for this, active observation of the guest OS and memory protection of the guest OS are proposed.

HyperSentry~\cite{Azab:2010:HES:1866307.1866313} is another research example, aimed at integrity verification of the hypervisor itself. In HyperSentry, using a dedicated observation mechanism and communication path called out-of-channel that cannot be detected by the verification target, the integrity of the hypervisor itself is verified. Concretely, a measurement mechanism is placed inside the hypervisor and communicates with a remote verifier via the SMI (System Management Interrupt) handler and IPMI (Intelligent Platform Management Interface). This realizes integrity verification of the entire system including the hypervisor, while securing sufficient isolation even from the hypervisor.

\subsection{Future Challenges}
\label{sec:virtualization_future}
Future challenges related to OS virtualization technology lie in further strengthening the various security functions related to virtualization technology by applying the research results introduced in the previous sections. This is because, while general-purpose OSs become ever larger and more complex and the possibility of their containing vulnerabilities increases, the tendency to adopt general-purpose OSs in various fields including embedded systems is expected to grow in the future; thus a mechanism that performs the construction and isolation of regions realizing security functions at the hypervisor---a layer closer to the hardware---is expected to become necessary.

For example, as strengthening of security functions related to access control, systems such as HUKO that detect and prevent unauthorized kernel extensions are currently being actively researched in many places. As for malware detection, study on hypervisors for malware detection and analysis is progressing, and its implementation method is currently a challenge. Research that detects malicious operations inside virtual machines at the hypervisor by resolving the semantic gap~\cite{Litty:2008:HSI:1496711.1496728} is also progressing.

\section{OS Verification Technology}
\label{sec:verification}
The OS is the software that forms the foundation of a computer system, and its reliability and safety greatly affect the security of the entire computer system. For this reason, research that directly verifies OS programs has been conducted for many years. In this section, we introduce three such methods: ``verification methods using theorem-proving assistants,'' ``verification methods using source-code model checking,'' and ``verification methods using safe programming languages.'' Table~\ref{tab:comparison_of_OS_verification_approaches} briefly compares the three verification methods. This section also touches on future challenges in OS verification.

\begin{table}
	\caption{Comparison of OS verification approaches}
	\centering
	\begin{tabular}{lcc}
		\toprule
		Verification method & Verifiable properties & Verification cost \\
		\midrule
		Theorem-proving assistants (Sec.~\ref{sec:proof_assistants}) & Excellent & Fair \\
		Source-code model checking (Sec.~\ref{sec:model_checking}) & Good & Good \\
		Safe programming languages (Sec.~\ref{sec:safe_languages}) & Fair & Excellent \\
		\bottomrule
	\end{tabular}
	\label{tab:comparison_of_OS_verification_approaches}
\end{table}

\subsection{Verification Methods Using Theorem-Proving Assistants}
\label{sec:proof_assistants}
A theorem-proving assistant is a program that takes as input some theorem and its proof and judges the correctness of that proof~\cite{nipkow2002,bertot2004}. First, the safety and reliability of the OS program are expressed as a theorem; next, its proof is created; and by verifying the correctness of that proof using the theorem-proving assistant, the safety and reliability of the OS are guaranteed and verified. Concretely, for example, by expressing the behavior of an OS program as state transitions on an abstract state machine and proving that the abstract state machine satisfies some property (for example, that the state machine does not perform illegal memory operations, or that it does not unexpectedly halt or fall into an infinite loop), the properties of the OS can be guaranteed and verified.

The advantage of verification methods using theorem-proving assistants is that any property that can be handled by the assistant can be verified. On the other hand, the disadvantage is that, because the proof of the theorem must be created by hand, the cost of verification becomes very large.

Kit~\cite{bevier1989} is the first OS kernel whose program was directly verified. Kit is a very small kernel consisting of about 300 lines of machine language (the machine language of a virtual von Neumann computer), and the independence of tasks within the kernel and the like are verified using the Boyer-Moore theorem prover~\cite{boyer1988}. Concretely, the abstract specification that the kernel should satisfy and the machine-language implementation are defined using Boyer-Moore logic, and it is proved using the Boyer-Moore theorem prover that the abstract specification guarantees the independence of each task within the kernel and that the implementation correctly realizes the abstract specification.

As an example of a larger-scale OS, seL4~\cite{klein2009} can be cited. seL4 is a kernel implemented with about 8,700 lines of C code and about 600 lines of assembly code, and that this implementation correctly realizes a formally given specification is guaranteed and verified using Isabelle/HOL~\cite{nipkow2002} (however, because seL4 is a microkernel, complex parts such as memory management lie outside the kernel and are not subject to verification). The actual verification work was carried out as follows. First, the properties and behaviors that the kernel should satisfy are defined as an abstract specification (for example, the interface and behavior of system calls). Next, as an abstract implementation of the abstract specification, an executable specification is defined. Concretely, a program implementing the abstract specification is written using a subset of the programming language Haskell~\cite{Haskell98Book}, and the executable specification is automatically generated from that source code. Then it is proved that the executable specification thus obtained correctly corresponds to the abstract specification. Next, the executable specification is implemented using a subset of the C language. Because the semantics of this C subset is formally defined in Isabelle/HOL, the C implementation can be handled almost directly in Isabelle/HOL. Finally, it is proved that this C implementation correctly corresponds to the executable specification. It is reported that, overall, these verification tasks required about 22 person-years.

\subsection{Verification Methods Using Source-Code Model Checking}
\label{sec:model_checking}
Source-code model checking~\cite{ball2001,henzinger2002} is a method that applies model-checking technology~\cite{clarke1999} directly to the source code of a program. More concretely, a model is extracted from the source code given as input, and by exhaustively checking the states that the program can take, the properties of the program are guaranteed and verified.

The advantage of verifying an OS using source-code model checking is that, unlike methods using theorem-proving assistants, it basically does not require human effort. On the other hand, a disadvantage is that, in general, model checking often requires a large amount of computational resources (CPU time, memory, etc.).

SDV (Static Driver Verifier)~\cite{ball2010} is a framework for verifying Windows device drivers. More concretely, whether a device driver follows the specification (mainly the usage of kernel APIs) written in a specification-description language called SLIC is verified using the source-code model checker SLAM~\cite{ball2004}. Verification of the OS kernel itself is not a target. SDV is included in Microsoft's device-driver development kit (Windows Driver Kit) and is already at a practical level.

Also, although source-code model checking is not directly applied, as an attempt to verify an OS kernel by a similar method, the verification of the part called the Nucleus in Verve~\cite{yang2010} can be cited. Verve is an OS whose type safety is guaranteed and verified; the Nucleus, which is a part of it, is mainly responsible for memory management (garbage collection), thread management (stack management), and hardware management (interrupt handling, device drivers, etc.). The verification of this Nucleus was carried out as follows. First, the assembly-code implementation of the Nucleus and the specification that the Nucleus should satisfy are expressed on a program verifier called Boogie~\cite{barnett2005} and passed to Boogie as input. Then Boogie generates verification conditions representing whether the Nucleus behaves according to the specification, and whether these conditions are satisfied is verified using an SMT solver called Z3~\cite{moura2008}. In this way, the verification of the Nucleus is basically automatic. The Nucleus consists of about 4,500 lines of Boogie code (this corresponds to about 1,400 lines of assembly code), of which about 7\,\% are annotations given by hand as hints for performing fully automatic verification. The description of the assembly-code implementation and specification in Boogie required about 9 person-months, and the automatic verification of the Nucleus by Boogie required 272 seconds on a machine with an Intel Core2 2.4\,GHz CPU and 4\,GB of memory.

\subsection{Verification Methods Using Safe Programming Languages}
\label{sec:safe_languages}
In this paper, ``safe programming language'' refers to a programming language that can guarantee and verify, through strict type checking and the like, that a program does not perform illegal memory operations, code execution, and so on at run time. By describing an OS program using such a safe programming language, it can be guaranteed and verified, through type checking and the like, that the OS does not perform illegal memory operations and the like. Strict type checking analyzes a program, classifies the variables and expressions in the program by type (a classification representing the kind of data or code, such as integer or memory reference), and checks whether the program behaves at run time according to that classification (type). For this reason, a program that passes strict type checking is guaranteed not to perform illegal memory operations or code execution that would cause type errors.

The advantage of describing an OS program using a safe programming language is that verification of the OS program is performed automatically by the type checking of the programming language. There is also the advantage that, compared with source-code model checking, the computational resources (CPU time, memory, etc.) required for verification are small. On the other hand, there is the disadvantage that the verifiable properties are limited to basic ones. There is also the problem that, for general OS developers, describing an OS using a safe programming language has a high barrier to entry.

SPIN~\cite{bershad1995} is a microkernel-style OS kernel that can safely extend its functions. In traditional microkernels, the safety of the kernel is ensured by executing kernel extensions at a privilege level different from that of the kernel body, but this had the problem that overhead arises in communication between the kernel extensions and the kernel, and in communication between kernel extensions. In contrast, the kernel extensions of SPIN can be written using the type-safe programming language Modula-3~\cite{modula3}, so their safety is guaranteed even when they are executed at the same privilege level as the kernel body, and the overhead due to communication can be reduced. Concretely, processing that requires low latency, such as virtual-memory-region management and network-communication processing, is implemented as kernel extensions using Modula-3. Note that in SPIN the kernel body itself is not subject to verification. Also, if there is a bug in the Modula-3 compiler, the safety of the kernel may be broken; that is, the Modula-3 compiler must be included in the system's Trusted Computing Base (TCB).

Singularity~\cite{hunt2005} is an OS written in the type-safe programming language C\#~\cite{csharp}, the programming language Sing\# (which extends C\#), and assembly language. Strictly speaking, Sing\# is a further extension of Spec\#, which itself is an extension of C\#. Spec\# extends C\# so that the preconditions and postconditions of object methods, and the like, can be explicitly described. Sing\# further extends Spec\# and introduces channels as a means of inter-process communication. In Singularity, each component of the OS is implemented as a process, and communication between components is performed through the aforementioned channels. Concretely, Singularity consists of about 280,000 lines of C\# program and about 90,000 lines of Sing\# (and architecture-dependent assembly programs), and the type safety of the parts written in C\# and Sing\#, the independence between components, and so on, are guaranteed. However, parts such as memory management and thread management are outside the direct verification target. Note that the C\# and Sing\# programs are converted into a kind of type-safe assembly program (typed assembly language~\cite{morrisett1999}), and type safety can be checked at the assembly-program level, so (unlike SPIN) there is no need to include the C\#/Sing\# compiler in the system's TCB.

TOS~\cite{tosh2009} is an OS kernel written using TALK~\cite{tosh2006,tosh2010}, a typed assembly language extended for describing OS kernels. TALK makes it possible to handle, in the type system, variable-length arrays for representing memory, integer constraints for handling address computation, and safe strong updates of memory (memory operations that change the type of a memory region), thereby enabling the description and type checking of OS kernels that were difficult with conventional typed assembly languages. TOS is a function-limited OS kernel consisting of about 3,000 lines of TALK program, but it is characterized by the fact that parts such as memory management and thread management---which were not directly targeted in other research on OS verification using safe programming languages---are also described in TALK, and their memory safety can be guaranteed and verified by TALK's type checking.

\subsection{Future Challenges}
\label{sec:verification_future}
As introduced up to the previous section, direct verification of OS programs has already been demonstrated at the research level, and the future challenge is to apply verification technology to larger-scale and more practical OSs. To realize this, it is considered realistic to combine and use several of the verification technologies introduced in the previous sections. In fact, in the aforementioned Verve~\cite{yang2010}, the parts other than the Nucleus are separately written in C\# (i.e., a safe programming language), so that the type safety of the OS kernel as a whole is guaranteed.

Also, as more academic and technical challenges, how to verify the safety of OS programs in environments where multiple programs are executed simultaneously---such as the multi-core CPUs that have become widespread in recent years~\cite{tosh2010}---can be cited. Conventional OS-verification research basically assumes a single-CPU computer environment; in particular, it becomes necessary to deal with race conditions that arise when multiple programs simultaneously operate on shared memory, and with problems of the memory consistency model~\cite{adve1996,boudol2009,atig2010}.

\section{Access Control Technology}
\label{sec:access_control}
Access control technology is one of the most basic element technologies supporting information security, and it relates to the protection of confidentiality, integrity, and availability alike. For this reason, as described in Section~\ref{sec:overview}, research that provides access-control functions from the OS as the TCB to upper layers and thereby guarantees information security has been continuously conducted since nearly half a century ago, and early research results can be seen in the TCSEC (Trusted Computer System Evaluation Criteria)~\cite{Orange:1983}. In recent years, a highly abstracted model of an access-control system has been standardized as ISO/IEC~10181-3~\cite{iso10181-3}. In the following subsections, as the main constituent elements of OS access-control functions---security policy models, security policy description languages, security policy verification technology, and access control mechanisms---we introduce recent research trends for each and describe future challenges.

\subsection{Security Policy Models}
\label{sec:smodel}
Because information systems require various levels of confidentiality, integrity, and availability according to their use, various security policy models reflecting these characteristics have been studied. The main categories are the confidentiality-protection type and the integrity-protection type, in addition to hybrid types of these; representative security policy models of each category include, as a confidentiality-protection type, the Bell-LaPadula model~\cite{mls}; as integrity-protection types, the Biba Integrity model~\cite{biba} and the Clark-Wilson model~\cite{clark1987comparison}; and as hybrid types, the Chinese Wall model~\cite{brewer1989chinese} and the RBAC model~\cite{ferraiolo1995role,ferraiolo1992role}. Below we show representative recent research examples of security policy models.

The RBAC model~\cite{ferraiolo1995role,ferraiolo1992role} is a security policy model centered on the Role; because it makes it easy to realize access control within an information system in accordance with real-world organizational structures and each person's responsibilities, it is intuitively easy to understand, and if operated correctly it can realize access control along the principle of least privilege. For these reasons, its adoption is advancing in various OSs, beginning with SELinux and Solaris. Recent research results on the RBAC model include the formal definition of the RBAC model and the extension of the basic model~\cite{sandhu1996role}, and the standardization of the RBAC model and its adoption into the ANSI standard~\cite{ferraiolo2001proposed}. In addition, the RBAC model has an administrative challenge of who should set, and how, the various relationships within an RBAC system; in response to this, there are the studies of ARBAC~\cite{sandhu1999role,sandhu1999arbac97} and an analysis of the RBAC model focusing on administrative challenges~\cite{crampton2003administrative}. Also, a method for formally verifying the security properties of an RBAC model implemented in an actual OS has been proposed~\cite{jha2007towards}.

The TBAC (Task-Based Authorization Controls) model~\cite{Thomas:1997:TAC:646115.679940} is a security policy model that groups multiple authorizations as an Authorization Step (AS) into an authorization procedure, and determines the authorization decision by specifying, in addition to the access subject, access target, and operation content, the AS name and the process name within the AS. In TBAC, the access rights granted to a subject for each AS are sequentially enabled and disabled while considering their order relation, so that, for example, fine-grained access rights can be specified according to the processing status of a transaction. Thereby, TBAC can abstractly structure a particular authorization procedure and make it reusable like a subroutine, thus improving the overall transparency of the policy.

\subsection{Security Policy Description Languages}
\label{sec:pl}
A security policy description language is for expressing security policies, and it can be seen as a language for describing the specification of access control. In recent years, research on security policy description languages has applied insights from mathematical logic such as predicate logic; the main challenges include expressive power for the delegation, restriction, and negation of access rights, grammatical clarity and readability for humans, concise and clear semantics, efficient authorization-decision procedures, and extensibility. Below we show representative recent research examples of security policy description languages.

SecPAL~\cite{Becker:1270639} is a security policy description language for flexibly describing the delegation of access rights between distributed administered domains, premised on configuring security policies modularized per administrative domain in relatively large-scale systems. SecPAL is a high-level security policy description language based on a constraint logic language; an authorization-decision request is approved when a query against a set of clauses succeeds. SecPAL's grammar is close to natural language, and its semantics consists of three inference rules; by supporting negative queries, recursive predicates, count-specifiable delegation of authority, and various other restrictions, it can generically express many security policy models. SecPAL is still being researched in order to be implemented as the security policy description language of the cloud OS Windows Azure.

Lithium~\cite{Lithium} is a security policy description language that can formally and correctly reason about logical negation. Lithium is a high-level language based on first-order predicate logic, and realizes reasoning that includes negation by restricting recursive expressions. Lithium is useful, for example, in that it can formally confirm that a certain access is unauthorized when merging multiple policies and analyzing their access-control specification; on the other hand, it has the issue that it cannot concisely express the delegation of access rights, which is needed in many security policy models. For Lithium, a front-end~\cite{rosetta} has also been developed that enables policy description in ordinary English grammar, so that it can be used without knowledge of its grammar or of first-order predicate logic.

\subsection{Security Policy Verification}
\label{sec:pv}
Security policy verification is a technology for verifying that a security policy satisfies a particular specification regarding access control. Because recent information systems are large-scale and complex, the amount of security-policy description by languages such as those introduced in the previous section tends to increase especially, and often the realized access-control specification cannot be clearly grasped by humans. To deal with this challenge, verification methods for easily analyzing described security policies are demanded, and active research is being conducted. Below we show representative recent research examples of security policy verification.

RBAC-PAT~\cite{Gofman2009} can verify, for the ARBAC model, the reachability of information flow starting from a certain Role, the availability, the compartmentalization of the information system, the weakest point, and so on. RBAC-PAT deals with the challenge that, in ARBAC models where changes to access rights are made at any time by multiple administrators, policy analysis becomes difficult. Concretely, it can verify the reachability and availability of a specific Role for a certain user, relations such as containment and minimal sets among Roles, the discovery of unnecessary Roles, and information flows between objects.

PALMS~\cite{guttman2005verifying} can verify all information flows existing in an MLS policy under analysis by defining a formal specification of the MLS policy. It can also automatically verify the consistency between two MLS policies. PALMS is implemented as a Prolog-based tool, and it can verify that a given MLS policy satisfies the $\ast$-property and the simple security condition, that an application's MLS policy is consistent with the MLS policy of its host OS, and so on. PALMS targets only MLS policies, but because in real information systems multiple security policy models often cooperate, research is progressing to enable verification when an MLS policy cooperates with other security policy models as well.

\subsection{Access Control Mechanisms}
\label{sec:mechanism}
An access control mechanism is a mechanism for enforcing on the information system the individual access-control rules described by a security policy description language. As concrete implementation methods, the ACL method and the Capability method have existed conventionally~\cite{Saltzer:1451869}; the ACL method has been widely adopted for a long time, but because the Capability method has the advantage of making it easy to realize access control based on the principle of least privilege, research on the Capability method has also continued. In addition, research on adding to existing OSs a mandatory access control mechanism that emphasizes cooperation with security policies is also being conducted. Below we show representative recent research examples of access control mechanisms.

seL4~\cite{klein2009} is a research OS that performs security extensions based on the L4 microkernel; it implements the take-grant model~\cite{Lipton:1977:LTA:322017.322025} combined with the Capability method, realizing access control over various kernel objects. Here, the kernel objects of L4 are threads, address spaces, inter-process communication, and the untyped-memory representing unused physical memory; access rights to these are granted and revoked through Capabilities. Another research example using the Capability method is Capsicum~\cite{Watson:2010:CPC:1929820.1929824}. Capsicum is a research OS that extends the standard UNIX API, and by using Capabilities it realizes sandboxing of processes and applications easily and at a fine granularity. Capsicum is being researched with the design philosophy that, rather than centrally controlling the compartmentalization of the application layer from the OS, the OS supports each individual application in easily realizing its own compartmentalization.

The Flask security architecture (FLSA)~\cite{Spencer99theflask} consists of an object manager that mandatorily enforces decisions made by the security policy, and a security server that makes access decisions according to the given security policy. Each performs data communication according to a predetermined protocol, and FLSA as a whole fulfills the role of a reference monitor. A major feature of FLSA is that it can flexibly realize various security policy models. SELinux is an implementation of FLSA for Linux; in addition, SEBSD is also an implementation example of FLSA. As a similar mandatory access control mechanism, the TrustedBSD MAC Framework~\cite{Watson:2001:TAT:647054.715753} has been proposed and is implemented in FreeBSD and Darwin.

\subsection{Future Challenges}
\label{sec:access_control_future}
Access control technology has been researched for a relatively long time, and its most basic theory has already been established. At the same time, in accordance with the changing demands on information security, its extension and application are continuously required, and in recent years, in line with the becoming-social-infrastructure of information systems, various studies such as those introduced in the previous sections are in progress.

Future challenges are to apply these results to environments where the required access-control specifications and the premises for realizing them differ, such as cloud environments based on virtualization technology and embedded environments such as smartphones. Concretely, security policy models for distributed systems, their description languages, and distributed access control mechanisms that enforce them are demanded, and research toward this is already being studied in many places. For example, access control mechanisms for extending the applicable scope of FLSA from targets that can be captured at the OS layer of a stand-alone system to targets in the application layer and other systems are being researched~\cite{SESQL,PMS}.

\section{Conclusion}
\label{sec:conclusion}
In this paper, we organized various recent research on the OS in relation to information security, and presented the challenges of each technology field. In organizing the research, we classified it from the correspondence with the design requirements of the reference monitor, focusing in particular on virtualization technology, OS verification technology, and access control technology; and after further subdividing each technology field, we described an overview, research examples, and challenges for each item. Concretely, we subdivided and explained virtualization technology into the observation of virtual machines by the hypervisor, the virtualization of main memory, the virtualization of the I/O mechanism, and the integrity verification of virtual machines; OS verification technology into verification methods using theorem-proving assistants, verification methods using source-code model checking, and verification methods using safe programming languages; and access control technology into security policy models, security policy description languages, security policy verification, and access control mechanisms.

For the sound development of the future information society, security support by a robust OS is indispensable; but, as introduced in this paper, the various element technologies for realizing this still hold many challenges. This is presumably because the demands on the OS did not change greatly for many years, and in particular, for OSs in practical use, their basic principles have remained almost unchanged for nearly half a century. However, looking ahead, in line with the rapid spread of new computing environments, the demands on the OS are also expected to change more than ever before; in particular, considering the importance of information security and the OS's suitability as the naturally-equipped TCB and reference monitor, it is hoped that the remaining challenges of the various technologies taken up in this paper will be solved, and that they will spread to be used as a matter of course in the real world. We would be glad if this paper became a foothold for those who will undertake this work in the near future, and a help in supporting the development of information security research in our country.

\section*{Acknowledgments}
We thank the members of the editorial committee and the anonymous reviewers for their valuable advice.

\bibliographystyle{unsrtnat}
\bibliography{refs}

\end{document}